\documentclass[manuscript,11pt]{aastex}
\usepackage{enumerate}

\input blackdvi.tex

\def\el61{EL$_{61}$}

\lefthead{Levison et al$.$}
\righthead{Origin of 2003~EL$_{61}$}

\begin{document}

\title{ON A SCATTERED-DISK ORIGIN FOR THE 2003~EL$_{61}$ COLLISIONAL
  FAMILY \Red{--- AN EXAMPLE OF THE IMPORTANCE OF COLLISIONS ON THE
    DYNAMICS OF SMALL BODIES}}

\author{Harold F$.$ Levison}
\affil{Southwest Research Institute\\ 
1050 Walnut St, Suite 300\\ 
Boulder, CO 80302}
\authoremail{hal@boulder.swri.edu}

\author{Alessandro Morbidelli}
\affil{Observatoire de la C\^ote d'Azur\\
  Boulevard de l'Observatoire\\
  B.P. 4229, 06304 Nice Cedex 4, France}

\author{David Vokrouhlick\'y}
\affil{Institute of Astronomy, Charles University\\
V Hole\v sovi\v ck\'ach 2\\
18000 Prague 8, Czech Republic}

\and

\author{William F$.$ Bottke}
\affil{Southwest Research Institute\\ 
1050 Walnut St, Suite 300\\ 
Boulder, CO 80302}

\newpage

\begin{abstract}
  
  The recent discovery of the 2003~EL$_{61}$ collisional family in the
  Kuiper belt (Brown et al$.$~2007) is surprising because the
  formation of such a family is a highly improbable event in today's
  belt.  Assuming Brown et al$.$'s estimate of the size of the
  progenitors, we find that the probability that a Kuiper belt object
  was involved in such a collision since primordial times is less than
  roughly $0.001$.  In addition, it is not possible for the collision
  to have occurred in a massive primordial Kuiper belt because the
  dynamical coherence of the family would not have survived whatever
  event produced the currently observed orbital excitation.  Here we
  suggest that the family is the result of a collision between two
  scattered disk objects.  We show that the probability that a
  collision occurred between two such objects with sizes similar to
  those advocated in Brown et al$.$~(2007) and that the center of mass
  of the resulting family is on an orbit typical of the Kuiper belt
  can be as large as 47\%.  Given the large uncertainties involved in
  this estimate, this result is consistent with the existence of one
  such family.  If true, this result has implications far beyond the
  origin of a single collisional family, because it shows that
  collisions played an important role in shaping the dynamical
  structure of the small body populations that we see today.

\end{abstract}


\newpage
\section{Introduction}
\label{sec:intro}

Recently, Brown et al$.$~(2007, hereafter BBRS) announced the
discovery of a collisional family associated with the Kuiper belt
object known as (136108) 2003~EL$_{61}$ (hereafter referred to as
\el61).  With a diameter of $\sim\!1500\,$km (Rabinowitz et
al$.$~2006), \el61\ is the third largest known Kuiper belt object.
The family so far consists of \el61\ plus seven other objects that
range from 150 to $400\,$km in diameter (BBRS; Ragozzine \&
Brown~2007\Red{, hereafter RB07}).  Their proper semi-major axes ($a$)
are spread over only $1.6\,$AU, their eccentricities ($e$) differ by
less than 0.08, and their inclinations ($i$) differ by less than
$1.5^\circ$ (see Table~\ref{tab:fam}). After correcting for some drift
in the eccentricity of \el61\ \Red{and 1999 OY$_3$ due to Neptune's
  mean motion resonances}, this corresponds to a velocity dispersion
of $\lesssim\!150\,$m/s (\Red{RB07}).  BBRS estimates that there is
only a one in a million chance that such a grouping of objects would
occur at random.

\begin{table}[h]
\singlespace
\begin{center}
\begin{tabular}{|lccc|}
\hline
name           &  $a$    &   $e$  &    $i$  \\
               &  (AU)   &        &   (deg) \\
\hline
2003 EL$_{61}$       &   43.3  & 0.19  &   28.2  \\
1995 SM$_{55}$       &   41.7  & 0.10  &   27.1  \\
1996 TO$_{66}$       &   43.2  & 0.12  &   27.5  \\
1999 OY$_3$     &   44.1  & 0.17  &   24.2 \\ 
2002 TX$_{300}$      &   43.2  & 0.12  &   25.9  \\
2003 OP$_{32}$       &   43.3  & 0.11  &   27.2  \\
2003 UZ$_{117}$ &   44.1  & 0.13  &   27.4 \\
2005 RR$_{43}$       &   43.1  & 0.14  &   28.5  \\
        \hline
\end{tabular}
\caption{\label{tab:fam} The orbital elements of the known \el61\
  family as supplied by the {\it Minor Planet Center} on July 20,
  2007.}
\end{center}
\end{table}

Based on the size and density of \el61\ and on the hydrodynamic
simulations of Benz \& Asphaug~(1999), and assuming an impact velocity
of $3\,$km/s, BBRS estimate that this family is the result of an
impact between two objects with diameters of $\sim\!1700\,$km and
$\sim\!1000\,$km.  Such a collision is surprising (M$.$~Brown, pers.
comm.), because there are so few objects this big in the Kuiper belt
that the probability of the collision occurring in the age of the
Solar System is very small.

Thus, in this paper we investigate the circumstances under which a
collision like the one needed to create the \el61\ family could have
occurred.  In particular, in \S{\ref{sec:KB}} we look again at the
idea that the larger of the two progenitors of this family (the
target) originally resided in the Kuiper belt and carefully determine
the probability that the impact could have occurred there.  We show
that this probability is small, and so we have to search for an
alternative idea.  In \S{\ref{sec:SD}}, we investigate a new scenario
where the \el61\ family formed as a result of a collision between two
scattered disk objects (hereafter SDOs).  The implications of these
calculations are discussed in \S{\ref{sec:end}}.

\section{The Kuiper belt as the source of the target}
\label{sec:KB}

In this section we evaluate the chances that a Kuiper belt object
roughly $1700\,$km in diameter could have been struck by a
$\sim\!1000\,$km body over the age of the Solar System.  There are two
plausible sources of the impactor: the Kuiper belt itself, and the
scattered disk (Duncan \& Levison~1997 hereafter DL97; see Gomes et
al$.$~2007 for a review). We evaluate each of these separately.

\subsection{The Kuiper belt as the source of the impactor}
\label{ssec:KBKB}

We start our discussion with an estimate of the likelihood that a
collision similar to the one described in BBRS could have
occurred between two Kuiper belt objects over the age of the Solar
System.  Formally, the probability ($p$) that an impact will occur
between between two members of a population in time $t_l$ is:
\begin{equation}
\label{eq:p}
p = N_i\, N_t\, t_l\, (R_i+R_t)^2\,\bar\varrho,
\end{equation}
where $N$ is the number of objects, $R$ is their radii, and the
subscripts $i$ and $t$ refer to the impactors and targets,
respectively.  In addition, $\bar\varrho$ is the {\it mean intrinsic}
impact rate which is the average of the probability that any two
members of the population in question will strike each other in any
given year assuming that they have a combined radius of $1\,$km. As
such, $\bar\varrho$ is only a function of the orbital element
distribution of the population.  For the remainder of this subsection
we append the subscript $kk$ to both $p$ and $\bar\varrho$ to indicate
that we are calculating these values for Kuiper belt --- Kuiper belt
collisions.

To evaluate Eq$.$~\ref{eq:p}, we first use the Bottke et al$.$~(1994)
algorithm to calculate the intrinsic impact rate between each pair of
orbits in a population. The average of these rates is
$\bar\varrho_{kk}$.  Using the currently known multi-opposition Kuiper
belt objects, we find that $\bar\varrho_{kk}\!=\!1.8 \times
10^{-22}\,{\rm km}^{-2}\,{\rm y}^{-1}$.  However, this distribution
suffers from significant observational biases, which could, in
principle, affect our estimate.  As a check, we apply this calculation
to the synthetic Kuiper belts resulting from the formation simulations
by Levison et al$.$~(2008).  These synthetic populations are clearly
not affected by observational biases, but may not represent the real
distribution very well. As such, although they suffer from their own
problems, these problems are entirely orthogonal to those of the
observed distribution.  We find $\bar\varrho_{kk}$'s between $1.5
\times 10^{-22}$ and $1.6 \times 10^{-22}\,{\rm km}^{-2}\,{\rm
  y}^{-1}$.  The fact that the models and observations give similar
results gives us confidence that our answer is accurate despite the
weaknesses of the datasets we used.  We adopt a value of
$\bar\varrho_{kk}\!=\!1.7 \times 10^{-22}\,{\rm km}^{-2}\,{\rm
  y}^{-1}$.

Next, we need to estimate $N_i$ and $N_t$.  Roughly 50\% of the sky
has been searched for Kuiper belt objects (KBOs) larger than
$1000\,$km in radius, and two have been found: Pluto and Eris (Brown
et al$.$~2005; Brown \& Schaller~2007).  Thus, given that almost all
of the ecliptic has been searched, let us assume that there are 3 such
objects in total.  Recent pencil beam surveys have found that the
cumulative size distribution of the Kuiper belt is $N(>\!R) \propto
R^{-3.8}$ for objects the size of interest here (Petit et al$.$~2006).
Thus, there are roughly 5 objects in the KBOs consistent with
BBRS's estimate of the size of the target body
($R_t\!=\!850\,$km) and $\sim\!40$ impactors ($R_i\!=\!500\,$km).
Plugging these numbers into Eq$.$~\ref{eq:p}, we find that the
probability that the impact that formed the \el61\ family could have
occurred in the current Kuiper belt is only $2.5\times 10^{-4}$ in the
age of the Solar System.
 
In the above discussion, we are assuming that the Kuiper belt has
always looked the way we see it today.  However, it (and the rest of
the trans-Neptunian region) most likely went through three distinct
phases of evolution (see Morbidelli et al$.$~2007 for a review):

\begin{enumerate}[1)]
  
\item At the earliest times, Kuiper belt objects had to have been in
  an environment where they could grow.  This implies that the disk
  had to have been massive (so that collisions were common) and
  dynamically quiescent (so that collisions were gentle and led to
  accretion; Stern~1996; Stern \& Colwell~1997a; Kenyon \& Luu~1998;
  1999).  Indeed, numerical experiments suggest that the disk needed
  to contain tens of Earth-masses of material and have eccentricities
  significantly less than 0.01 (see Kenyon et al$.$~2008 for a
  review).  In what follows we refer to this quiescent period as {\it
    Stage~I}.
  
\item The Kuiper belt that we see today is neither massive nor
  dynamically quiescent.  The average eccentricity of the Kuiper belt
  is $\sim\!0.14$ and estimates of its total mass range from
  0.01~$M_\oplus$ (Bernstein et al$.$~2004) to 0.1~$M_\oplus$ (Gladman
  et al$.$~2001).  Thus, there must have been some sort of dynamical
  event that significantly excited the orbits of the KBOs.  This event
  was either violent enough to perturb $>\!99\%$ of the primordial
  objects onto planet-crossing orbits thereby directly leading to the
  Kuiper belt's small mass (Morbidelli \& Valsecchi~1997; Nagasaki \&
  Ida$.$~2000; Levison \& Morbidelli~2003; Levison et al$.$~2008), or
  excited the Kuiper belt enough that collisions became erosional
  (Stern \& Colwell~1997b; Davis \& Farinella~1997; Kenyon \&
  Bromley~2002, 2004).  It was during this violent period that most of
  the structure of the Kuiper belt was established.  As we discuss
  below, the Kuiper belt's resonant populations might be the only
  exception to this.  Indeed, the inclination distribution in the
  resonances shows that these populations either formed during this
  period or post-date it (Hahn \& Malhotra~2005).  It is difficult to
  date this event.  However, there has been some work that suggests
  that it might be associated with the late heavy bombardment of the
  Moon, which occurred 3.9 Gy ago (Levison et al$.$~2008).  In what
  follows we refer to this violent period as {\it Stage~II}.

\item Since this dramatic event, the Kuiper belt has been relatively
  quiet.  Indeed, the only significant dynamical changes may have
  resulted from \Red{the gradual decay of intrinsically unstable
    populations and} the slow outward migration of Neptune.  As
  discussed in more detail in \S{\ref{sec:SD}}, this migration occurs
  as a result of a massive scattered disk that formed during Stage~II.
  It might be responsible for creating at least some of the resonant
  structure seen in the Kuiper belt (Malhotra~1995; Hahn \&
  Malhotra~2005).  This migration continues today, although at an
  extremely slow rate.  We refer to this modern period of Kuiper belt
  evolution as {\it Stage~III}.

\end{enumerate}

Perhaps the simplest way to resolve the problem of the low probability
of an \el61-like collision is to consider whether this event could
have occurred during Stage~I, when the Kuiper belt may have been 2 to
3 orders of magnitude more populous than today (see Morbidelli et
al$.$~2007 for a review).  Increasing the $N$'s in Eq$.$~\ref{eq:p} by
a factor of 100--1000 would not only make a collision like the one
needed to make the \el61\ family much more likely, but it would make
them ubiquitous. Indeed, this explains why many large KBOs (Pluto and
Eris, for example) have what appear to be impact-generated satellites
(e$.$g$.$ Canup~2005; Brown et al$.$~2005).

However, the fact the we see the \el61\ family in a tight clump in
orbital element space implies that if the collision occurred during
Stage~I, then whatever mechanism molded the final structure of the
Kuiper belt during Stage~II must have left the clump intact.  Three
general scenarios have been proposed to explain the Kuiper belt's
small mass: {\it (i)} the Kuiper belt was originally massive, but the
strong dynamical event in Stage~II caused the ejection of most of the
bodies from the Kuiper belt to the Neptune-crossing region (Morbidelli
\& Valsecchi~1997; Nagasawa \& Ida~2000), {\it (ii)} the Kuiper belt
was originally massive, but the dynamical excitation in Stage~II
caused collisions to become erosive\footnote{Recall that by
  definition, most of the collisions that occurred during Stage~I were
  accretional.}  and thus most of the original Kuiper belt mass was
ground to dust (Stern \& Colwell~1997b; Davis \& Farinella~1997;
Kenyon \& Bromley~2002, 2004), and {\it (iii)} the observed KBOs
accreted closer to the Sun, and during Stage~II a small fraction of
them were transported outward and trapped in the Kuiper belt by the
dynamical evolution of the outer planets (Levison \& Morbidelli~2003;
Levison et al$.$~2008).

Scenario~{\it (ii)} cannot remove objects as large as the \el61\ 
precursors because the collisions are not energetic enough.  Indeed,
in order to get this mechanism to explain the Kuiper belt's small
mass, almost all of the original mass must have been in objects with
radii less then $\sim\!10\,$km (Kenyon \& Bromley~2004).  Thus, in
this scenario, the number of $\sim\!500\,$km objects present at early
epochs is no different than what is currently observed.  Thus, this
scenario cannot solve our problem.  Scenarios~{\it (i)} and {\it
  (iii)} invoke the wholesale dynamical transport of most of the
Kuiper belt.  While, this can remove most of the targets and
impactors, the dynamical shakeup of the Kuiper belt would obviously
destroy the coherence of the family.  This is due to the fact that any
dynamical mechanism that could cause such an upheaval would cause the
orbits of the KBOs to become wildly chaotic, and thus any tight clump
of objects would spread exponentially in time.  From these
considerations we conclude that the collision that created the \el61\ 
family could not have occurred between two KBOs (see Morbidelli~2007
for further discussion).

\subsection{The scattered disk as the source of the impactor}
\label{ssec:KBSD}

In this section we evaluate the probability that the larger progenitor
of the \el61\ family originally was found in the Kuiper belt, but the
impactor was a member of the scattered disk.  For reasons described
above, the family-forming impact must have occurred some time during
Stage~{III} when the main dynamical structure of the Kuiper belt was
already in place.  However, the scattered disk is composed of
trans-Neptunian objects that have perihelia near enough to Neptune's
orbit that their orbits are not stable over the age of the Solar
System (see Gomes~2007 for a review).  As a result, they are part of a
dynamically active population where objects are slowly diffusing
through orbital element space and occasionally leave the scattered
disk by either being ejected from the Solar System, evolving into the
Oort cloud, or being handed inward by Neptune, thereby becoming
Centaurs.

Therefore, unlike the Kuiper belt, the population of the scattered
disk has slowly been decreasing since its formation and this decay is
ongoing even today.  It is an ancient structure (Morbidelli et
al$.$~2004; Duncan et al$.$~2004) that was probably constructed during
Stage~II, and thus has slowly evolved and decayed in number during all
of Stage~III.  DL97 estimated that the primordial scattered
disk\footnote{In what follows, when we refer to the `primordial
  scattered disk' we mean the scattered disk that existed at the end
  of Stage~II and at the beginning of Stage~III.}  may have contained
roughly 100 times more material at the beginning of Stage~III than we
see today.  We need to include this evolution in our estimate of the
collision probability.

The above requirement forces us to modify Eq$.$~\ref{eq:p}.  In
particular, since we have to assume that the number of Kuiper belt
targets ($N_t$) has not significantly changed since the beginning of
Stage~III (Duncan et al$.$~1995),
\begin{equation}
\label{eq:pt}
p_{sk} = (R_i+R_t)^2\,\bar\varrho_{sk}\,N_t\, \int{N_i(t) dt},
\end{equation}
where the subscript $sk$ refers to the fact that we are calculating
these values for SDO---KBO collisions.  \Red{In writing
  Eq.~\ref{eq:pt} in the manner, we are assuming that the scattered
  disk orbital element distribution, and thus $\bar\varrho_{sk}$, does
  not significantly change with time.  In all the calculations
  discussed below, we find that this is an accurate assumption.}

Assuming that the size distribution of SDOs does not change with time,
we can define $f(t) \equiv N_{i}(t)/N_{i0}$, where $N_{i0}$ is the
number of impactors at the beginning of Stage~III.  As a result,
$\int{N_i(t) dt}$ in Eq$.$~\ref{eq:pt} becomes $N_{i0}\int{f\,dt}$.
Now, if we define $\bar{t} \equiv \int{f\,dt}$, then $p_{sk}$ takes on
the same form as in Eq$.$~\ref{eq:p}:
\begin{equation}
\label{eq:ptt2}
p_{sk} = N_{i0}\, N_{t}\, \bar{t}\, (R_i+R_t)^2\,\bar\varrho_{sk}.
\end{equation}

\begin{figure}[h!]
\vglue 3.0truein
%\special{psfile=PS/noft.ps angle=0 hoffset=120 voffset=-50 vscale=40 hscale=40}
\includegraphics{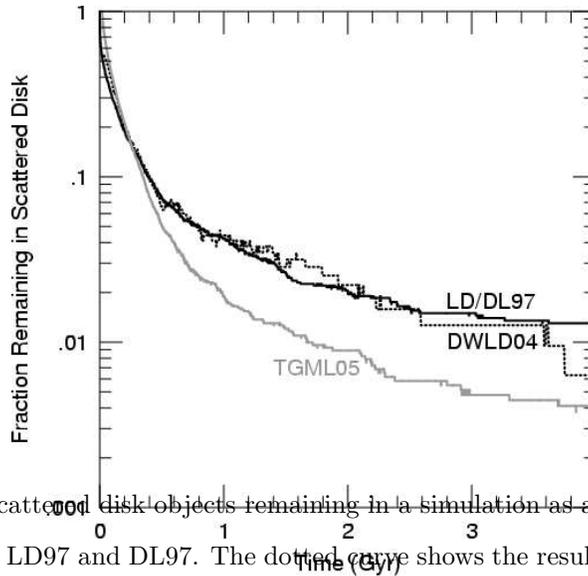}
\caption{\footnotesize \label{fig:noft}
  {The fraction of scattered disk objects remaining in a simulation as
    a function of time.  The solid curve shows the results from LD97
    and DL97.  The dotted curve shows the results from DWLD04.  The
    gray curve shows TGML05.  Time is measured from the beginning of
    Stage~III.}}
\end{figure}

We must rely on dynamical simulations in order to estimate $f(t)$ and
$\bar{t}$.  In addition, our knowledge of the orbital element
distribution (and thus $\varrho_{sk}$) of SDOs is hampered by
observational biases on a scale that is much worse than exists for the
Kuiper belt because of the larger semi-major axes involved.  Thus, we
are required to use dynamical models to estimate $\varrho_{sk}$ as
well.  For this purpose, we employ three previously published models
of the evolution of the scattered disk:

\begin{enumerate}
  
\item {\it LD/DL97:} Levison \& Duncan~(1997) and DL97 studied the
  evolution of a scattered disk whose members originated in the Kuiper
  belt.  In particular, they performed numerical orbital integrations
  of massless particles as they evolved from Neptune-encountering
  orbits in the Kuiper belt.  The initial orbits for these particles
  were chosen from a previous set of integrations whose test bodies
  were initially placed on low-eccentricity, low-inclination orbits in
  the Kuiper belt but then evolved onto Neptune-crossing orbits
  (Duncan et al$.$~1995).  The solid curve in Figure~\ref{fig:noft}
  shows the relative number of SDOs as a function of time in this
  simulation.  After $4\times 10^{9}\,$yr, 1.25\% of the particles
  remain in the scattered disk.  We refer to this fraction as $f_s$
  (see Table~\ref{tab:val}). Note that $f_s$ is equivalent to
  $f(4Gy)$.  In addition, we find that $\bar{t}=1.9\times 10^8\,$y in
  this integration.
  
\item {\it DWLD04:} Dones et al$.$~(2004) studied the formation of the
  Oort cloud and dynamical evolution of the scattered disk from a
  population of massless test particles initially spread from 4 to
  $40\,$AU with a surface density proportional to $r^{-3/2}$.  For the
  run employed here, the \Red{initial} RMS eccentricity and
  inclination were 0.2 and $5.7^\circ$, respectively.  Also, we
  restricted ourselves to use only those objects with initial
  perihelion distances $<\!32\,$AU.  The dotted curve in
  Figure~\ref{fig:noft} shows the relative number of SDOs as a
  function of time in this simulation.  For this model $f_s = 0.63\%$
  and $\bar{t}=3.9\times 10^8\,$y.
  
\item {\it TGML05:} Tsiganis et al$.$~(2005, hereafter TGML05)
  proposed a new comprehensive scenario --- now often called `the Nice
  model' --- that reproduces, for the first time, many of the
  characteristics of the outer Solar System.  It quantitatively
  recreates the orbital architecture of the giant planet system
  (orbital separations, eccentricities, inclinations; Tsiganis et
  al$.$~2005).  It also explains the origin the Trojan populations of
  Jupiter (Morbidelli et al$.$~2005) and Neptune (Tsiganis et
  al$.$~2005; Sheppard \& Trujillo~2006), and the irregular satellites
  of the giant planets (Nesvorn\'y et al$.$ 2007a).  Additionally, the
  planetary evolution that is described in this model can be
  responsible for the early Stage~II evolution of the Kuiper belt
  (Levison et al$.$~2008).  Indeed, it reproduces many of the Kuiper
  belt's characteristics for the first time.  It also naturally
  supplies a trigger for the so-called Late Heavy Bombardment (LHB) of
  the terrestrial planets that occurred $\sim\!3.9$ billion years ago
  (Gomes et al$.$~2005).

\medskip

TGML05 envisions that the giant planets all formed within
$\sim\!15\,$AU of the Sun, while the known KBOs formed in a massive
disk that extended from just beyond the orbits of the giant planets to
$\sim\!30\,$AU.  A global instability in the orbits of the giant
planets led to a violent phase of close planetary encounters
(Stage~II).  This, in turn, caused to Uranus and Neptune to be
scattered into the massive disk.  Gravitational interactions between
the disk and the planets caused the dispersal of the disk (some
objects being pushed into the Kuiper belt; Levison et al$.$~2008) and
forced the planets to evolve onto their current orbits (see also
Thommes et al$.$~1999; 2002).  After this violent phase (i$.$e$.$ at
the beginning of Stage~III), the scattered disk is massive.  As in the
other models above, it subsequently decays slowly due to the
gravitational effects of Neptune.  The gray curve in
Figure~\ref{fig:noft} shows the relative number of SDOs as a function
of time in TGML05's nominal simulation\footnote{TGML05 stopped their
  integrations at 348 Myr.  Here we continued their simulation to
  $4\,$Gyr using the RMVS integrator (Levison \& Duncan 1994),
  assuming that the disk particles were massless.}.  We set $t=0$ to
be the point at which the orbits of Uranus and Neptune no longer
cross.  For this model $f_s = 0.41\%$ and $\bar{t}=1.5\times 10^8\,$y.

\end{enumerate}

Once the $f$'s are known, all we need in order to calculate
Eq$.$~\ref{eq:ptt2} is $N_{i0}$, which, recall, is the initial number
of $1000\,$km diameter impactors in the scattered disk, and
$\varrho_{sk}$.  To evaluate $N_{i0}$, we need to combine our
dynamical models with observational estimates of the scattered disk.
The most complete analysis of this kind to date is by Trujillo et
al$.$~(2000).  These authors performed a survey of a small area of the
sky in which they discovered three scattered disk objects.  They
combined these data with those of previous surveys, information about
their sky coverage, limiting magnitudes, and dynamical models of the
structure of the scattered disk to calculate the number of SDOs with
radii larger than $50\,$km.  To perform this calculation, they needed,
however, to assume a size distribution for the scattered disk.  In
particular, they adopted $N(>\!R) \propto R^{-q}$, and studied cases
with $q\!=\!2$ and 3.

Unfortunately, if we are to adopt Trujillo et al$.$'s estimates of the
number of SDOs, we must also adopt their size distributions, because
the former is dependent on the latter.  This might be perceived to be
a problem because we employed a much steeper size distribution for the
Kuiper belt in \S{\ref{ssec:KBKB}}.  Fortunately, $q\!=\!3$ is in
accordance with the available observations for this population.  In
particular, it is in agreement with the most modern estimate of
Bernstein et al$.$~(2004), who found $q=3.3^{+0.7}_{-0.4}$ for bright
objects (as these are) in a volume limited sample of what they call
the `excited class' (which includes the scattered disk).  In addition,
it is consistent with the results of Morbidelli et al$.$~(2004), who
found that $2.5\,\lesssim\,q\,\lesssim\,3.5$ for the scattered disk.
Also note that Bernstein et al$.$~(2004) concluded that the size
distribution of their `excited class' is different from the rest of
the Kuiper belt at the 96\% confidence level, which again supports the
choices we make here.  Thus, we adopt Trujillo et al$.$'s estimate for
$q\!=\!3$, which is that there are between 18,000 and 50,000 SDOs with
$R\!>\!50\,$km and $50\!<\!a\!<\!200\,$AU.  We also adopt $q\!=\!3$ in
the remainder of this discussion.\footnote{Note that if we had adopted
  $q\!=\!3$ in \S{\ref{ssec:KBKB}}, our final estimate of the
  probability that the \el61\ family was the result of a collision
  between two KBOs ($p_{kk}$) would have actually been {\it smaller}
  by about a factor of two. This, therefore, would strengthen the
  basic result of this paper.}

LD/DL97's model of the scattered disk places about 66\% of its SDOs in
Trujillo et al$.$'s range of semi-major axes.  This fraction is 47\%
in DWLD04 and 40\% in TGML05.  Thus, we estimate that there are
currently between 27,000 and 125,000 SDOs larger than $50\,$km
($N_{\rm 50km}$) depending on the model.  So, the initial number of
objects in the scattered disk of radius $R$ is
\begin{equation}
\label{eq:n0}
N_{0}(R) = \frac{N_{\rm 50km}}{f_s} \left(\frac{R}{50{\rm km}}\right)^{-q}.
\end{equation}
The values of $N_{i0}$ derived from this equation are presented in
Table~\ref{tab:val} for Trujillo et al$.$'s value of $q=3$.

\begin{table}[h]
\singlespace
\begin{center}
\begin{tabular}{|l|ccc|}
\hline
               &               LD/DL97    &  DWLD04  &    TGML05  \\
\hline
$f_s$          &   1.3\%          &  0.63\%            &    0.41\%  \\
$\bar{t}$ (y)     & $1.9\times 10^8$ &  $3.9\times 10^8$ & $1.5\times 10^8$ \\ 
$\bar\varrho_{sk}$ (km$^{-2}$y$^{-1}$)  
           & $\Red{6.7}\times 10^{-23}$ & $\Red{7.2}\times 10^{-23}$ & $\Red{1.1 \times 10^{-22}}$ \\
$N_{i0}$       & 2180 -- 6060     &  6980 -- 16900    & 11,000 --- 30,400 \\
$p_{sk}$ &  $\Red{2.4}\times 10^{-4}$ -- $\Red{6.9}\times 10^{-4}$ &
               $\Red{1.7}\times 10^{-3}$ -- $\Red{4.3}\times 10^{-3}$    & $\Red{1.6}\times 10^{-3}$ -- $\Red{4.5}\times 10^{-3}$ \\

\hline
$\Delta V_{\rm min}$ (m/s) & \Red{198}    &   \Red{263}         &  \Red{93} \\
$t_l$ (y)      & $3.4\times 10^7$ &  $1.5\times 10^8$ & $4.6\times 10^7$ \\ 
$N_{t0}$       &  440 -- 1230     & 1240 -- 3440      & 2230 -- 6260 \\
$\bar\varrho_{ss}$ (km$^{-2}$y$^{-1}$)  
           & $1.1\times 10^{-22}$ & $8.0\times 10^{-23}$ & $7.9\times 10^{-23}$ \\
$p_{ss}$         &  0.007 -- 0.051     &  0.16  -- 1.27     & 0.16 -- 1.26 \\
$p_{\rm KB}$  &    0.19              & 0.076             & 0.32 \\
$p_{\rm SD}$  & $1.5\times 10^{-3}$ -- 0.011 & 0.012 --
               0.10  & 0.061 -- 0.47 \\
\hline
\end{tabular}
\caption{\label{tab:val}  Important dynamical parameters derived from
  the three pre-existing scattered disk models.  See text for a full
  description.}
\end{center}
\end{table}

Finally, we need $\bar\varrho_{sk}$, which, recall, only depends on
the orbital element distribution of the targets and impactors.  We can
take the orbital element distribution of the impactors directly from
our scattered disk numerical models, but we need to assume the orbit
of the target.  \Red{We place the target on the center of mass orbit
  for the family as determined by RB07.  This orbit has
  $a\!=\!43.1\,$AU, $e\!=\!0.12$, and $i\!=\!28.2^\circ$.}  As before,
the values of $\bar\varrho_{sk}$ are calculated using the Bottke et
al$.$~(1994) algorithm and are also given in the table.  \Red{It was
  somewhat surprising to us that the values for $\bar\varrho_{sk}$ are
  so similar to $\bar\varrho_{kk}$ because the scattered disk is
  usually thought of as a much more extended structure.  However, we
  found the median semi-major axis of objects in our scattered disk
  simulations is only about $60\,$AU.  This is similar enough to the
  Kuiper belt to explain the similarity.}

It should be noted that \Red{the Bottke et al$.$} algorithm assumes a
uniform distribution of orbital angles, which might be of some doubt
for the scattered disk.  As a result, we tested these distributions
for our objects with semi-major axes between 40 and $200\,$AU and
found that, although there was a slight preference for arguments of
perihelion near 0 and 180$^\circ$, the distributions were uniform to
better than one part in ten.

We can now evaluate $p_{sk}$ for the various models.  These too are
given in Table~\ref{tab:val}. We find that the probability that the
\el61\ family is the result of a collision between a Kuiper belt
target with a radius of $850\,$km and a scattered-disk impactor with a
radius of $500\,$km is less than 1 in 220.  Although this number is
larger than that for Kuiper belt -- Kuiper belt collisions, it is
still small.  Thus, we conclude that we can rule out that idea that
the progenitor (i$.$e$.$ the target) of the \el61\ family was in the
Kuiper belt.

\section{The scattered disk as the source of both the target and impactor}
\label{sec:SD}

In the last section we found that an SDO-KBO collision is much more
likely than a KBO-KBO collision because the scattered disk was more
massive in the past.  Thus, in order to increase the overall
probability of a \el61\ family forming event even further, we need to
investigate whether {\it both} the target and the impactor could have
been in the scattered disk at the time of the collision. This
configuration has the advantage of increasing the number of potential
targets by roughly 2 orders of magnitude relative to the estimate
employed in \S{\ref{ssec:KBSD}}, at least at the beginning of
Stage~III. At first sight, the assumption that both progenitors were
in the scattered disk may seem at odds with the fact that the family
is found in the Kuiper belt today.  Remember, however, that collisions
preserve the total linear momentum of the target and the impactor.  As
a result, the family is dispersed around the center of mass of the two
colliding bodies, {\bf not} around the orbit of the target.  If the
relative velocity of the colliding objects is comparable to their
orbital velocity and the two bodies have comparable masses, then the
center-of-mass of the resulting family can be on a very different
orbit than the progenitors.
  
With this in mind, we propose that at some time near the beginning of
Stage~III, two big scattered disk objects collided.  Before the
collision, each of them was on an eccentric orbit typical of the
scattered disk.  At the time of the collision, one object was moving
inward while the other was moving outward, so that the center of mass
of the target-projectile pair had an orbit typical of a Kuiper belt
object. As a result, we should find the family clustered around this
orbit today.

We start our investigation of the above hypothesis by determining
whether it is possible for the center of mass of two colliding SDOs to
have a Kuiper belt orbit like that of the \el61\ family.  We
accomplish this by comparing $\Delta V_{\rm min}$ to $\delta V_{\rm
  min}$, where $\Delta V_{\rm min}$ is defined to be the \Red{minimum}
difference in velocity between the \el61 family orbit and the
scattered disk region, and $\delta V_{\rm min}$ is the possible
difference in velocity between the center of mass of the collision and
the original orbit of the target.  \Red{If $\Delta V_{\rm
    min}\!>\!\delta V_{\rm min}$ then a collision between two SDOs
  cannot lead to \el61's orbit. If, on the other hand, $\Delta V_{\rm
    min}\!<\!\delta V_{\rm min}$, our scenario is at least possible.
  Note, however, that this condition is necessary, but not sufficient,
  because the orientation of the impact is also important.  This
  effect will be accurately taken into account in the numerical models
  performed later in this section.}

\begin{figure}[h!]
\vglue 7.5truein
\includegraphics{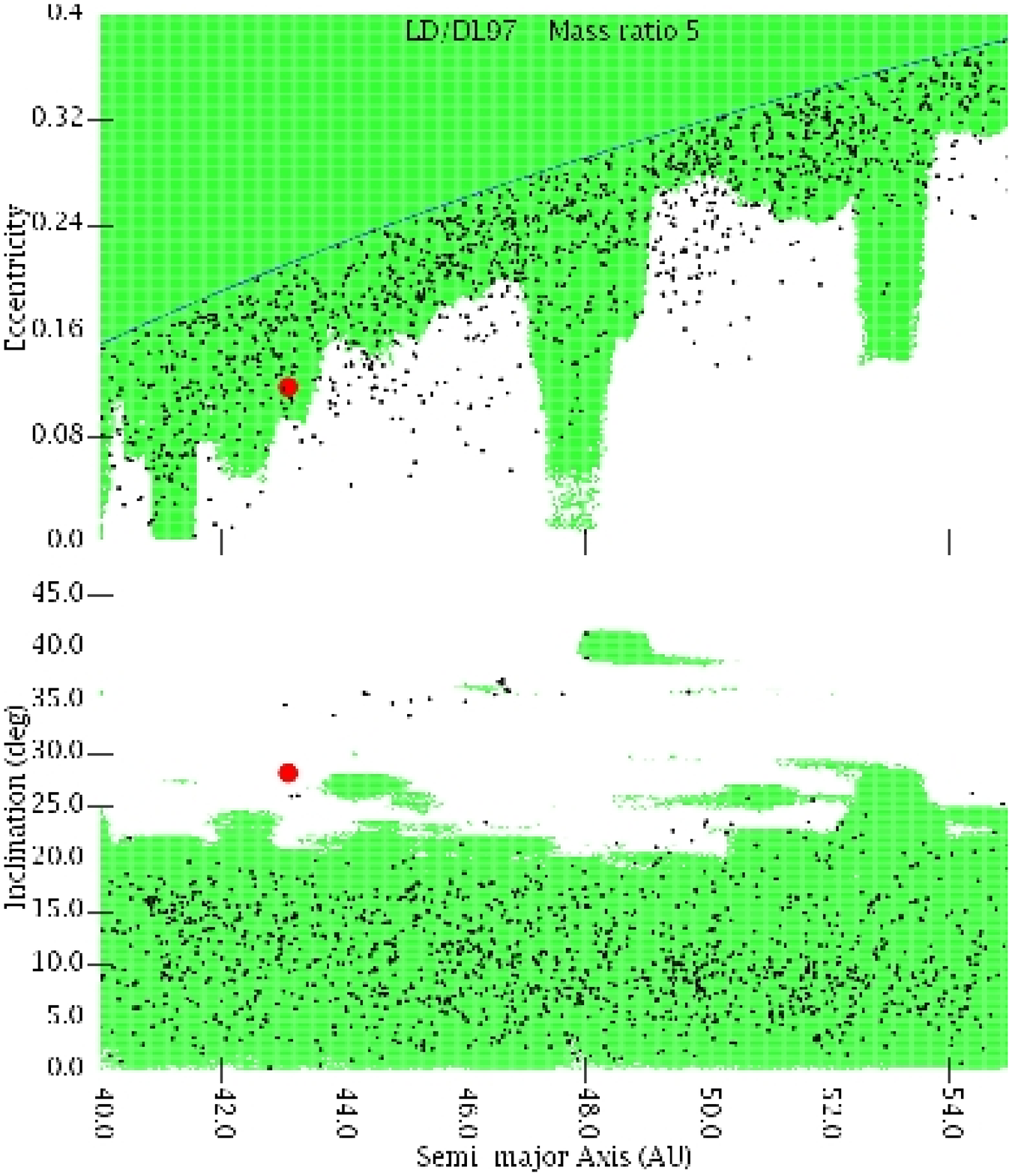}
\includegraphics{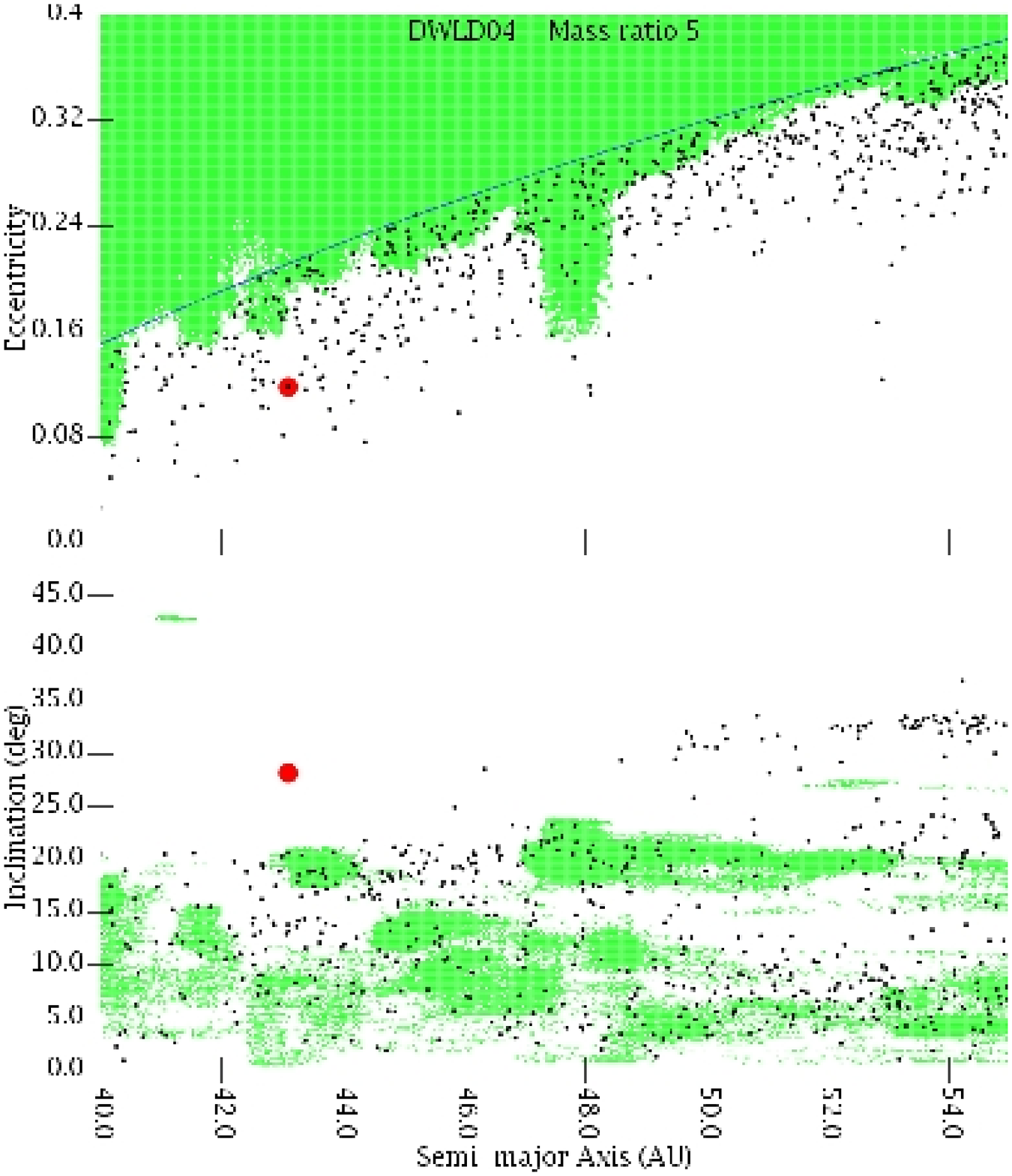}
\includegraphics{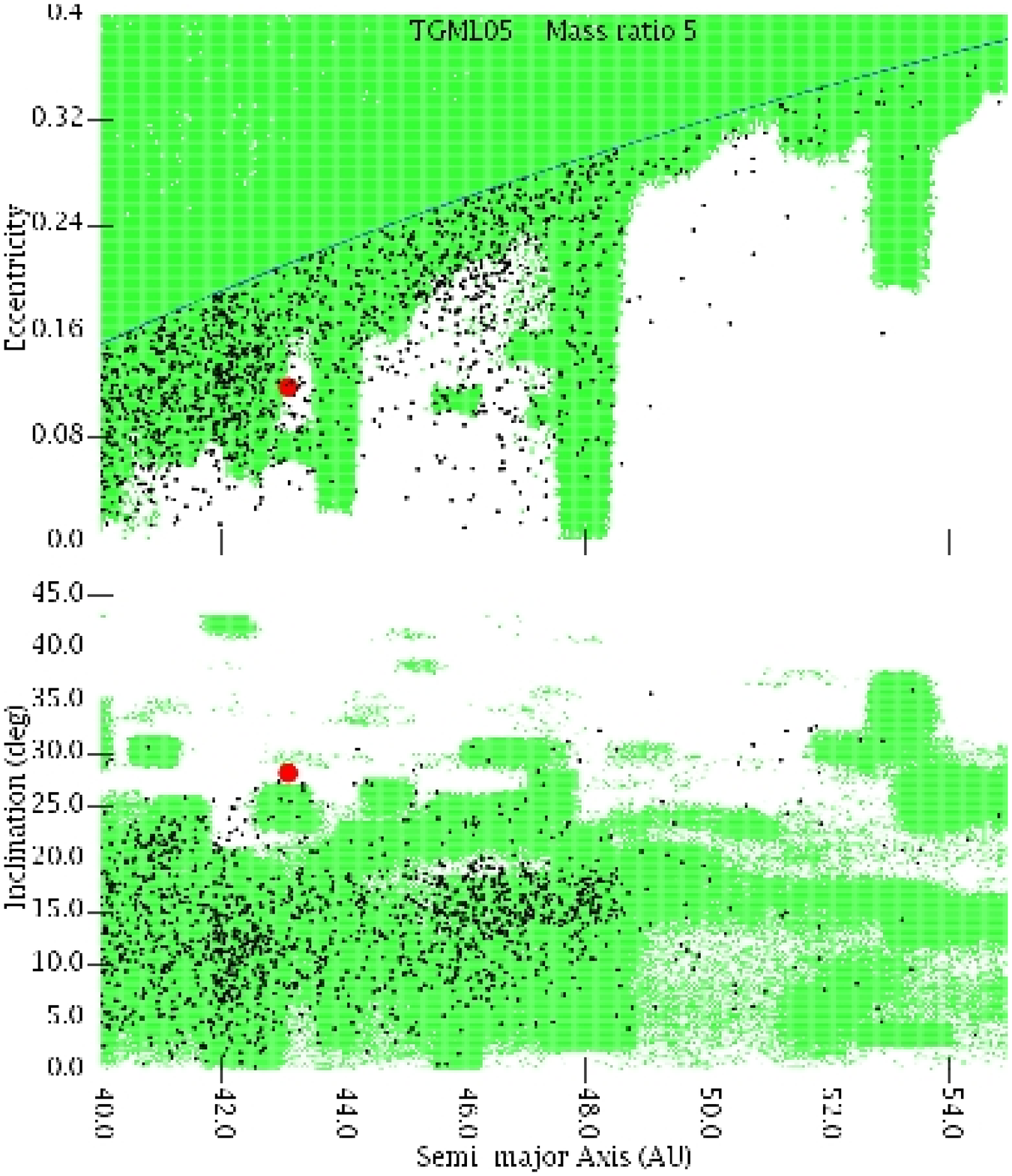}

%\special{psfile=PS/aei_LD97.ps angle=0 hoffset=-80 voffset=200 vscale=60 hscale=60}
%\special{psfile=PS/aei_DWLD.ps angle=0 hoffset=170 voffset=200 vscale=60 hscale=60}
%\special{psfile=PS/aei_Nice.ps angle=0 hoffset=45 voffset=-100 vscale=60 hscale=60}
\caption{\footnotesize \label{fig:aei}
  {The green area illustrates the regions (eccentricity --- semi-major
    axis distribution on the top panels and inclination --- semi-major
    axis distribution on the bottom panels) visited by scattered disk
    objects with no collisions included (left panels: from LD/DL97;
    right panels: from DWLD04; bottom TGML05).  The blue curve in the
    top panels marks $q\!=\!34\,$AU and the \Red{red dot shows the
      center of mass orbit of the \el61\ family (RB07).}  The
    \Red{minimum} difference in orbital velocity between the \el61\ 
    family and the scattered disk visited region is $\Red{265}\,$m/s.
    For reference, a typical impact of scattered disk bodies with a
    mass ratio of 5 (as for the target/impactor estimated in BBRS)
    gives a $\delta V_{\rm min}$ of $\sim 450$~m/s.  The black dots
    show stable Kuiper belt orbits that result from actual simulations
    of SD evolution, accounting for such collisions.  In all cases the
    osculating orbits are shown.}}
\end{figure}

We start \Red{our simple comparision} with $\Delta V_{\rm min}$.  The
green areas in Figure~\ref{fig:aei} show the regions of orbital
element space visited by SDOs during our three $N$-body simulations.
It is important to note that these are two-dimensional projections of
the six-dimensional distribution consisting of all the orbital
elements.  Therefore, the fact that an area of one of the plots is
green does not imply that all the orbits that project into that region
belong to the scattered disk, only that some of them do.  The red
\Red{dot represents RB07's center of mass orbit for} the \el61\ 
family.  Note that the family is close to the region visited by SDOs.

The distance in velocity space between the location of the family and
the scattered disk region, $\Delta V_{\rm min}$, can be computed using
the techniques developed in Nesvorn{\' y} et al$.$~(2007b).  Given two
crossing orbits this algorithm uses Gauss' equations to seek the
minimum relative velocity ($\Delta V$) needed to move an object from
one orbit to another.  In particular, it searches through all values
of true longitudes and orbital orientations in space to find the
smallest $\Delta V$ while holding $a$, $e$, and $i$ of each orbit
fixed.  Using this algorithm, we take each entry from the orbital
distributions saved during the scattered disk $N$-body simulations and
compare it to \Red{RB07's center of mass orbit for} the \el61\ family.
We then take $\Delta V_{\rm min}$ to be the \Red{minimum} difference
in orbital velocity between the \el61\ family and the region visited
by scattered disk particles during our simulations.  These values are
listed in Table~\ref{tab:val}, and we find that all the $N$-body
simulations have particles which get within $\Red{265}\,$m/s of the
family.

Next we estimate $\delta V_{\rm min}$.  The center of mass velocity,
$\vec{V}_{\rm CM}$, of \Red{target-impactor system} is
$(m_i\vec{V}_i+m_t\vec{V}_t)/(m_i+m_t)$, where $m$ is the mass of
\Red{each body}.  So, $\delta V \equiv \vert\vec{V}_{\rm CM} -
\vec{V}_t\vert = \vert\vec{V}_i-\vec{V}_t\vert/(1+\frac{m_t}{m_i})$,
where $\vert\vec{V}_i-\vec{V}_t\vert$ is the impact speed, which BBRS
argues is roughly $3\,$km/s (in the simulations below we find the
average to be about $2.7\,$km/s). Therefore, assuming a mass ratio
between the target and impactor of 5 (as argued by BBRS), we expect
that the center of mass velocity (from which the fragments are
ejected) to be offset from the initial velocity of the target by about
$450\,$m/s.  Since this is larger than the minimum velocity distance
that separates the scattered disk from the \el61\ family ($\Delta
V_{\rm min}$; $<\!\Red{265}\,$m/s, as discussed above), it is possible
that the observed orbit of the \el61\ family could result from such a
collision.

We now estimate the likelihood that such a collision will happen.  To
accomplish this, we divide the problem into two parts.  We first
evaluate the probability ($p_{ss}$) that a collision occurred in the
age of the Solar System between two SDOs with $R_i\!=\!500\,$km and
$R_t\!=\!850\,$km.  Then, we calculate the likelihood ($p_{KB}$) that
the center of mass of the two colliding bodies was on a stable Kuiper
belt orbit.  Since fragments of the collision will be centered on this
orbit, the family members should span it.  In what follows, we refer
to this \Red{theoretical} orbit as the {\it `collision orbit'}.  The
probability that the \el61\ family originated in the scattered disk is
thus $p_{SD} = p_{ss} \times p_{KB}$.

As with the determination of $p_{sk}$ in \S{\ref{ssec:KBSD}}, we need
to modify Eq$.$~\ref{eq:p} to take into account that the number of
objects in the scattered disk is changing with time.  In this case,
however, both the number of targets and the number of impactors vary.
As result,
\begin{equation}
\label{eq:ptt}
p_{ss} = (R_i+R_t)^2\,\bar\varrho_{ss}\, \int{N_i(t)\,
  N_t(t)\, dt}. 
\end{equation}
Assuming that the size distribution of SDOs does not change with time,
$N_t(t)/N_{t0} = N_i(t)/N_{i0} = f(t)$, where $f(t)$ was defined
above.  Thus, $\int{N_i(t)\, N_t(t)\, dt}$ becomes
$N_{i0}N_{t0}\int{f^2\,dt}$.  Now, if we define $t_l \equiv
\int{f^2dt}$, then $p_{ss}$ again takes on the same form as in
Eq$.$~\ref{eq:p}:
\begin{equation}
\label{eq:pt2}
p_{ss} = N_{i0}\, N_{t0}\, t_l\, (R_i+R_t)^2\,\bar\varrho_{ss},
\end{equation}
where the subscript $ss$ refers to the fact that we are calculating
these values for SDO---SDO collisions. Note that $t_l$ is not the same
as $\bar{t}$ used in Eq$.$~\ref{eq:ptt2}, but it is a measure of the
characteristic time of the collision.  The values of $t_l$ for our
three scattered disk models are given in Table~\ref{tab:val}.

The values of $N_{i0}$ are the same as we calculated in
\S{\ref{ssec:KBSD}} using Eq$.$~\ref{eq:n0} because in both cases the
impacting population is the same.  In this case, we can also use
Eq$.$~\ref{eq:n0} to estimate $N_{t0}$.  These values are given in
Table~\ref{tab:val}.  The table also shows the values of
$\bar\varrho_{ss}$ for each of the models, which were again calculated
using the Bottke et al$.$~(1994) algorithm.  Recall that this
parameter only depends on the orbital element distribution of the
scattered disk.

We can now evaluate $p_{ss}$ for the various models.  These too are
given in Table~\ref{tab:val}. Again, we are assuming a target radius
of $850\,$km and a impactor radius of $500\,$km.  We find that our
scenario is least likely in the LD/DL97 model, with $p_{ss} \lesssim
0.06$, while it is most likely in the TGML05 model with
$p_{ss}\!\sim\!1$.  The fact that $p_{ss}$ can be close to one is
encouraging.  After all, we see one family and there are probably not
many more in this size range.  However, we urge caution in
interpreting these $p_{ss}$ values because there are significant
uncertainties in several of the numbers used to calculate them ---
particularly the $N$'s.  Indeed, we believe that the differences
between the $p_{ss}$ values from the various models are probably more
a result of the intrinsic uncertainties in our procedures rather than
the merit of one model over another.

Next, we need to calculate the probability that the impacts described
above have collision orbits in the Kuiper belt ($p_{KB}$). We
accomplish this with the use of a Monte Carlo simulation where we take
the output of our three orbital integrations and randomly choose
particle pairs to collide with one another based on their location and
the local number density.  We apply the following procedures to the
LD/DL97, DWLD04, and TGML05 datasets, separately.

Our preexisting $N$-body simulations supply us with a series of {\it
  snapshots} of the evolving scattered disk as a function of time.  In
particular, the original $N$-body code recorded the position and
velocity of each object in the system at fixed time intervals.  For
two objects to collide, they must be at the same place at the same
time.  However, because of the small number of particles in our
simulations (compared to the real scattered disk) and the fact that
the time intervals between snapshots are long, it is very unlikely to
find any actual collisions in our list of snapshots.  Thus, we must
bin our data in both space and time in order to generate pairs of
particles to collide.  For this purpose, we divided the Solar System
into a spatial grid. We assumed that the spatial distribution of
particles is both axisymmetric and symmetric about the mid-plane.
Thus, our grid covers the upper part of the meridional plane.  The
cylindrical radius ($\varpi$) was divided into 300 bins between 30 and
$930\,$AU, while the positive part of the vertical coordinate was
divided into 100 bins with $z\!\leq\!100\,$AU.  We also binned time.
However, since the number of particles in the $N$-body simulations
decreases with time (see Figure~\ref{fig:noft}), we increased the
width of the time bins ($\Delta t_{\rm bin}$) at later times in order
to insure we had enough particles in each bin to collide with one
another. In particular, we choose the width of each time bin so that
the total number of particles in the bin (summing over the spatial
bins) is the same.

We assigned each entry (meaning a particular particle at a particular
time) in the dataset of our original $N$-body simulation to a bin in
the 3-dimensional space discussed above (i$.$e$.$ $\varpi$--$z$--$t$).
As a result, the entries associated with each bin represent a list of
objects that were roughly at the same location at roughly the same
time in the $N$-body simulation.

Finally, we generated collisions at random.  This was accomplished by
first randomly choosing a bin based on the local collision rate, as
determined by a particle-in-the-box calculation.  It is important to
note that since the bins were populated using the $N$-body
simulations, this choice is consistent with the collision rates used
to calculate the mean collision probability $p_{ss}$ above.  As a
result, we are justified multiplying $p_{ss}$ and $p_{KB}$ together at
the end of this process.  Once we had chosen a bin, we randomly chose
a target and impactor from the list of objects in that bin.  From the
velocities of the colliding pair we determined the orbit of the pair's
center-of-mass assuming a mass ratio of 5.

The next issue is to determine whether these collision orbits are in
the Kuiper belt.  For this purpose, we define a KBO as an object on a
stable (for at least a long period of time) orbit with a perihelion
distance, $q$, greater than $34\,$AU (indicated by the blue curves in
Figure~\ref{fig:aei}).  To test stability, we performed a $50\,$Myr
integration of the orbit under the gravitational influence of the Sun
and the four giant planets.  As previous studies of the stability of
KBOs have shown (Duncan et al$.$~1995; Kuchner et al$.$~2002), a
time-span of $50\,$Myr adequately separates the stable from the
unstable regions of the Kuiper belt.  Any object that evolved onto an
orbit with $q\!<\!33\,$AU during this period of time was assumed to be
unstable.  The remainder were assumed to be stable and are shown as
the black dots in Figure~\ref{fig:aei}.

We find that collisions can effectively fill the Kuiper belt out to
near Neptune's 1:2 mean motion resonance at $48\,$AU.  We created
stable, non-resonant objects with $q$'s as large as $46.5\,$AU.
Indeed, the object with the largest $q$ has $a\!=\!47.3\,$AU,
$e\!=\!0.017$, and $i\!=\!18.2^\circ$ and thus it is fairly typical of
the KBOs that we see.  With regard to the \el61\ family, we easily
reproduce stable orbits with the same $a$ and $e$.  However, we find
that it is difficult to reproduce the family's inclination.  Although
we do produce a few orbits with inclinations larger than the family's,
$\sim\!90\%$ of the orbits in our simulations have inclinations less
than that of the \el61\ family.  

The lack of high inclination objects is clearly a limitation of our
model.  We believe, however, that this mismatch is more the result of
limitations in our scattered disk models than of our collisional
mechanism for capture in the Kuiper belt.  Neither the LD/DL97,
DWLD04, nor TGML05 simulations produce high enough inclinations to
explain what we see in the scattered disk.  So, if we had a more
realistic scattered disk model, we would probably be able to produce
more objects with inclinations like \el61\ and it cohorts.  One
concern of such a solution is that the higher inclinations would
affect our collision probabilities, particularly through their effects
on $\varrho_{ss}$.  To check this, we performed a new set of
calculations where we arbitrarily increased the inclinations of the
scattered disk particles by a factor of 2.  We find that the increased
inclinations decrease $\varrho_{ss}$ by less than 20\%.  Thus, we
conclude that if we had access to a scattered disk model with more
realistic inclinations, we should be able to better reproduce the
orbit of the family without significantly affecting the probability of
producing it.

The values of $p_{\rm KB}$ (the fraction of {\el61}-forming impacts
that lead to objects that are trapped in the Kuiper belt) resulting
from our main Monte Carlo simulations are listed in
Table~\ref{tab:val}.  Combining $p_{\rm KB}$ and $p_{ss}$ we find that
the probability that, in the age of the Solar System, two SDOs with
radii of $500\,$km and $850\,$km hit one another leading to a family
in the Kuiper belt (which we called $p_{\rm SD}$) is between 0.1\% and
47\%, depending on the assumptions we use.  For comparison, in
\S{\ref{sec:KB}} we computed that the probability that the \el61\ 
family is the result of the collision between two Kuiper belt objects
is $\sim\!0.02\%$, or is the result of a KBO--SDO collision is
$\lesssim\!0.1\%$\footnote{In \S{\ref{ssec:KBSD}}, we did not take
  into account the fact that collisions between a larger KBO and a
  somewhat smaller SDO could result in a family on an unstable orbit,
  i$.$e$.$ on an orbit that is not in the Kuiper belt.  Applying the
  above procedures to the collisions described in \S{\ref{ssec:KBSD}},
  we find that there is only a 29\% chance that the resulting family
  would be on a stable Kuiper belt orbit.  The values of $p_{sk}$ in
  Table~\ref{tab:val} should be multiplied by this factor.}. Thus, we
conclude that the progenitors of the \el61\ family are much more
likely to have originated in the scattered disk than in the Kuiper
belt.

\Red{Up to this point, we have been concentrating on whether our model
  can reproduce the observed center of mass orbit of the \el61\ 
  family.  However, the spread of orbits could also represent an
  important observational constraint (Morbidelli et al$.$~1995).  In
  particular, assuming that the ejection velocities of the collision
  were isotropic around the center of mass, the family members should
  fall inside an ellipse in $a$--$e$ and $a$--$i$ space.  The
  orientation and axis ratio of the ellipse in $a$--$e$ space are
  strong functions of the mean anomaly of the collision orbit at the
  time of the impact ($M$), while the axis ratio of the ellipse in
  $a$--$i$ space is a function of both $M$ and the argument of
  perihelion ($\omega$).  The major axis of the ellipse in $a$--$i$
  space should always be parallel to the $a$ axis}\footnote{\Red{This
    is indeed observed for the \el61\ family.  This fact strongly
    supports the idea the these objects really are the result of a
    collision and not simply a statistical fluke.}}.  \Red{Using this
  information, RB07 estimated that at the time of the collision, the
  center of mass orbit had $M\!=\!76^\circ$ and
  $\omega\!=\!271^\circ$.}

\Red{Given that we are arguing that the target and impactor originated
  in the scattered disk, we might expect that certain impact
  geometries are preferred, while others are forbidden.  Thus, we
  examined the $M$ and $\omega$ of all the collisions shown in
  Figure~\ref{fig:aei} (black dots) with orbits near that of RB07's
  center-of-mass orbit.  In particular, we chose collision orbits with
  $42\!<\!a\!<\!44\,$AU, $0.08\!<\!e\!<\!0.14$, and $i\!>\!15^\circ$.
  We found that we cannot constrain the values of $\omega$.  Indeed,
  these orbits are roughly uniform in this angle.  However, our model
  avoids values of $M$ between $-37^\circ$ and $62^\circ$, i$.$e$.$
  near perihelion.  This is a result of the fact that the collision
  must conserve momentum, lose energy, and that the initial orbits of
  the progenitors were in the scattered disk \Red{while the family
    must end up} in the Kuiper belt.  RB07's value of $M$ falls in the
  range covered by our models.}

\begin{figure}[h!]
\vglue 7.3truein
%\special{psfile=PS/orb_el.ps angle=0 hoffset=-50 voffset=-80 vscale=90 hscale=90}
\includegraphics{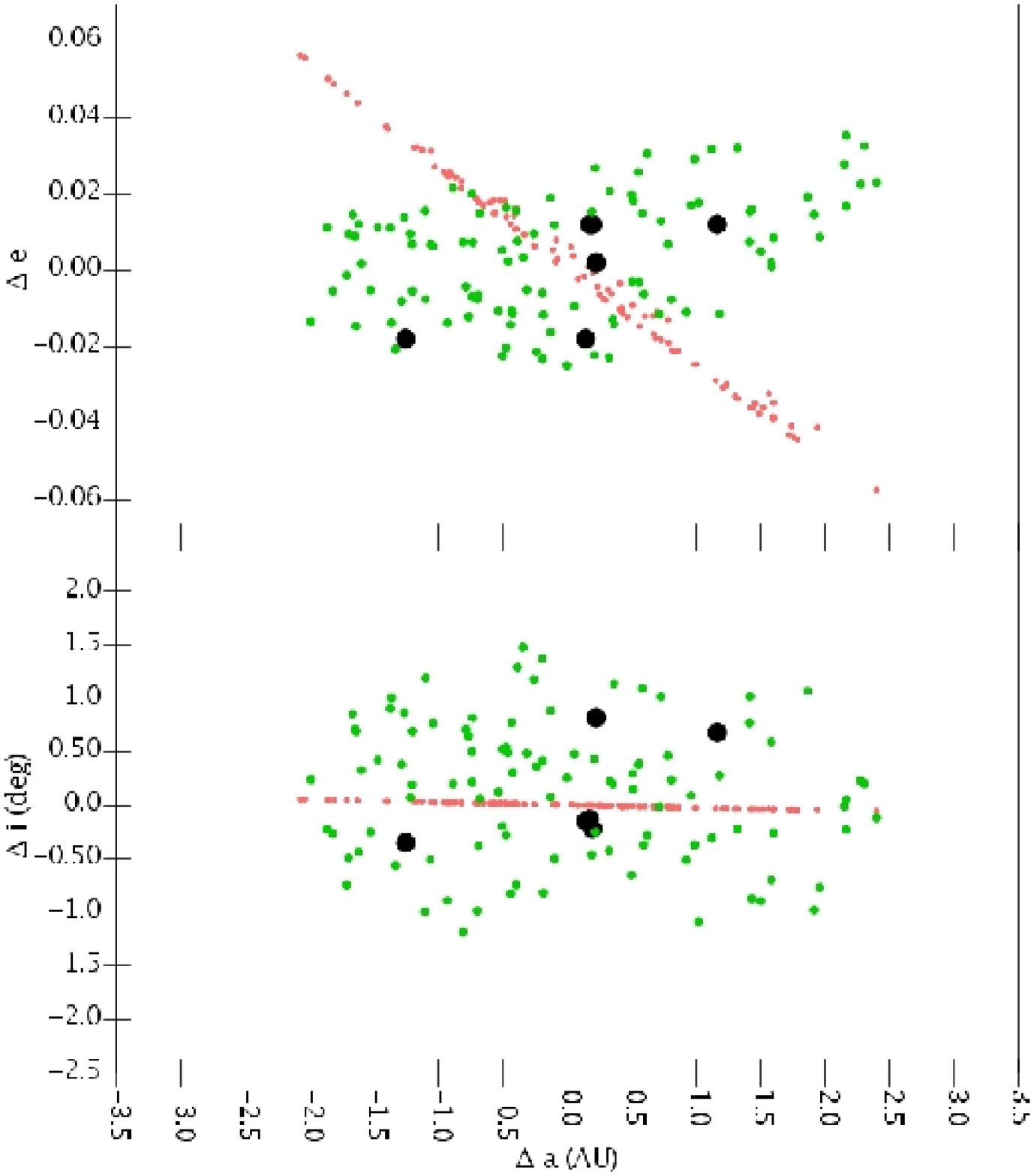}
\caption{\footnotesize \label{fig:orbel}
  { \Red{ A comparison of the spread of families in $\Delta
      a$--$\Delta e$ and $\Delta a$--$\Delta i$ space, where $\Delta
      x$ is defined to be the difference between a particular orbital
      element of the family member and that of the collision orbit.
      The black dots show the proper orbital elements of the real
      family members as determined by RB07.  We did not plot
      2003~EL$_{61}$ or 1999~OY$_{3}$ because their orbits have
      changed since the family formed (RB07).  The green dots show a
      fictitious family with a collision orbit of $a\!=\!42.0\,$AU,
      $e\!=\!0.09$, $i\!=\!21^\circ$, $\omega\!=\!111^\circ$,
      $M\!=\!-72^\circ$.  For comparison, the red dots show a
      fictitious family with a collision orbit of $a\!=\!42.2\,$AU,
      $e\!=\!0.09$, $i\!=\!23^\circ$, $\omega\!=\!294^\circ$,
      $M\!=\!174^\circ$.  This shows that this diagnostic is a
      sensitive test for the models and that we can reproduce the
      observations.}}}
\end{figure}

\Red{Figure~\ref{fig:orbel} shows a comparison between the spread of
  the \el61\ family in orbital element space (black dots) and two of
  our fictitious families.  The fictitious family members were
  generated by isotropically ejecting particles from the point of
  impact with a velocity of 150$\,$m/s (BBRS). The collision orbits
  for these families are consistent with the center of mass orbit for
  the family.  The collision orbit of the family shown in green has
  $M\!=\!71^\circ$ and $\omega\!=\!273^\circ$ --- similar to the
  values inferred by RB07.  For comparison, the family shown in red
  has an orbit with similar $a$, $e$, and $i$, but $M\!=\!174^\circ$
  and $\omega\!=\!294^\circ$.  We can conclude that, although this
  test is not very constraining because our model can reproduce most
  values and $M$ and $\omega$, we can match what is seen.}

There is one more issue we must consider.  In \S{\ref{ssec:KBKB}}, we
described the three phases of Kuiper belt evolution: 1) A quiescent
phase of growth (Stage~I), 2) a violent phase of dynamical excitation
and, perhaps, mass depletion (Stage~II), and 3) a relatively benign
modern phase (Stage~III).  We argued that any collisional family that
formed during Stage~I or Stage~II would have been dispersed during the
chaotic events that excited the orbits in the Kuiper belt.  Thus, the
family forming impact must have occurred during Stage~III.  However,
the fact that the violent evolution is over before the collision does
not mean that the orbits of the planets must have remained unchanged.
As a matter of fact, the decay of the scattered disk population
actually causes Neptune's orbit to slowly migrate outward.  This, in
turn, causes resonances to sweep through the Kuiper belt, potentially
affecting the orbits of some KBOs. So, as a final step in our
analysis, we must determine whether the dynamical coherence of the
\el61\ family would be preserved during this migration.

To address the above issue, we performed an integration of 100
fictitious family members on orbits initially with the same $a$, $e$,
and $i$ as \Red{RB07's center of mass orbit}, under the gravitational
influence of the four giant planets as they migrate.  We adopted the
case presented in Malhotra~(1995), where Neptune migrated from 23 to
$30\,$AU.  Note that the model in Levison et al$.$~(2008) has Neptune
migrating from $\sim\!27\,$AU, so we are adopting an extreme range of
migration here.  We found that only 12\% of the family members were
trapped in and pushed outward by Neptune's mean motion resonances
(i$.$e$.$ they were removed from the family).  The orbits of the
remaining particles were only slightly perturbed and thus they
remained recognizable family members.  Thus, we conclude that the
family would have survived the migration and that the SDO-SDO
collision is still a valid model for the origins of the \el61\ family.
Interestingly, however, this simulation predicts that we might find
family members (which can be identified by their IR spectra; BBRS) in
the more distant Neptune resonances (1:2, 2:5...).  If so, the
location of these objects can be used to constrain Neptune location at
the time when the \el61\ family formed.

So, we conclude that the most probable scenario for the origin of the
\el61\ family is that it resulted from a collision between two SDOs.
If true, this result has implications far beyond the origin of a
single collisional family because it shows, for the first time, that
collisions can affect the dynamical evolution of the Kuiper belt, in
particular, and small body populations, in general.  Indeed, this
process might be especially important for the so-called `hot'
classical Kuiper belt.  Brown~(2001) argued that the de-biased
inclination distribution of the classical Kuiper belt is bi-modal and
can be fitted with two Gaussian functions, one with a standard
deviation $\sigma \sim 2^\circ$ (the low-inclination {\it `cold'}
core), and the other with $\sigma \sim 12^\circ$ (the high-inclination
{\it `hot'} population).  Since the work of Brown, it has been shown
that the members of these two populations have different physical
properties (Tegler \& Romanishin~2000; Levison \& Stern~2001;
Doressoundiram et al$.$~2001), implying different origins.

Gomes~(2003) suggested that one way to explain the differences between
the hot and cold populations is that the hot population originated in
the scattered disk, because a small fraction of the scattered disk
could be captured into the Kuiper belt due to the gravitational
effects of planets as they migrated.  Here we show that collisions can
accomplish the same result.  Indeed, a collisional origin for these
objects may have the advantage of explaining why binaries with equal
mass-components are rarer in this population than in other parts of
the trans-Neptunian region.  Using HST, Noll et al$.$~(\Red{2008})
found that 29\% of classical Kuiper belt objects (see their paper for
a precise definition) with inclination $<\!5.5^\circ$ are similar-mass
binary objects, while this fraction is only 2\% for objects with
larger inclinations.  A collisional origin for the hot population
might explain this discrepancy because a collision that is violent
enough to kick an object from the scattered disk to the Kuiper belt
would also disrupt the binary (the binary member that was not struck
would have continued in the scattered disk).

One might expect that if the majority of the hot population was put in
place by collisions, we should be able to predict a relationship
between the size distribution of its members and that of the scattered
disk.  Eq$.$~\ref{eq:pt2} shows that the collision probability scales
roughly as $N^2$. And since in the scattered disk, $N(R) \propto
R^{-q}$, we might predict that the size distribution of the hot
population is $N_h(R) \propto R^{2-2q}$ (one power of $-q$ from both
$N_{i0}$ and $N_{0t}$, and a power of 2 from $(R_i+R_t)^2$, see
Eq$.$~\ref{eq:pt2}).  In this case $q\!\sim\!3$ (see above) and thus
$N_h(R) \sim R^{-4}$.  However, this estimate does not take into
account the fact that the collisions themselves could affect the size
distribution of the resulting hot population.  Unfortunately, it is
not yet clear what the size of the fragments would be because of poor
understanding of the collisional physics of icy objects at these
energies.  As a result, it is not yet possible to investigate this
intriguing idea.

\section{Conclusions}
\label{sec:end}

The recent discovery of the \el61\ family in the Kuiper belt
(BBRS) is surprising because its formation is, at first
glance, a highly improbable event.  BBRS argues that this
family is the result of a collision between two objects with radii of
$\sim\!850\,$km and $\sim\!500\,$km.  The chances that such event
would have occurred in the current Kuiper belt in the age of the Solar
System is roughly 1 in $4000$ (see \S{\ref{sec:KB}}).  In addition, it
is not possible for the collision to have occurred in a massive
primordial Kuiper belt because the dynamical coherence of the family
would not have survived whatever event molded the final Kuiper belt
structure.  We also investigated the idea that the family could be the
result of a target KBO being struck by a SDO projectile, and found
that the probability of such an event forming a family on a stable
Kuiper belt orbit is $\lesssim\!10^{-3}$.

In this paper, we argue that the \el61\ family is the result of a
collision between two scattered disk objects.  In particular, we
present the novel idea that the collision between two SDOs on highly
eccentric unstable orbits could damp enough orbital energy so that the
family members would end up on stable Kuiper belt orbits.  This idea
of using the scattered disk as the source of both of the family's
progenitors has the advantage of significantly increasing the
probability of a collision because the population of the scattered
disk \Red{was much larger in the early Solar System (it is currently
  eroding away due to the gravitational influence of Neptune --- DL97;
  DWLD04)}.  With the use of three pre-existing models of the
dynamical evolution of the scattered disk (DWLD04, LD/DL97, and
TGML05) we show that the probability that a collision between a
$\sim\!850\,$km SDO and $\sim\!500\,$km SDO occurred and that the
resulting collisional family was spread around a stable Kuiper belt
orbit can be as large as 47\%.  Given the uncertainties involved, this
can be considered on the order of unity.  Thus, we conclude that the
\el61\ family progenitors are significantly more likely to have
originated in the scattered disk than the Kuiper belt.

If true, this result has important implications for the origin of the
Kuiper belt because it is the first direct indication that collisions
can affect the dynamical evolution of this region.  Indeed, we
\Red{suggest} at the end of \S{\ref{sec:SD}} that this process might
be responsible for the emplacement of the so-called `hot' classical
belt (Brown~2001) because it naturally explains why so few of these
objects are found to be binaries (Noll et al$.$~\Red{2008}).

\acknowledgments HFL is grateful for funding from NASA's Origins, OPR,
and PGG programs.  AM acknowledges funding from the french National
Programme of Planetaology (PNP).  DV acknowledges funding from
\Red{funding from Grant Agency of the Czech Republic (grant
  205/08/0064) and the Research Program MSM0021620860 of the Czech
  Ministry of Education.  WB's contribution was paid for by the NASA's
  PGG, Origins, and NSF's Planetary Astronomy programs.  We would also
  like to thank Mike Brown, Luke Dones, Darin Ragozzine, and Paul
  Weissman for comments on early versions of this manuscript.}

\section*{References}

\begin{itemize}
\setlength{\itemindent}{-30pt}
\setlength{\labelwidth}{0pt}

\item[] Benz, W., Asphaug, E.\ 1999.\ Catastrophic Disruptions
  Revisited.\ Icarus 142, 5-20.

\item[] Bernstein, G.~M., Trilling, D.~E., Allen, R.~L., Brown, M.~E.,
  Holman, M., Malhotra, R.\ 2004.\ The Size Distribution of
  Trans-Neptunian Bodies.\ Astronomical Journal 128, 1364-1390.

\item[] Bottke, W.~F., Nolan, M.~C., Greenberg, R., Kolvoord, R.~A.\ 
  1994.\ Velocity distributions among colliding asteroids.\ Icarus
  107, 255-268.

\item[] Brown, M.~E.\ 2001.\ The Inclination Distribution of the
  Kuiper Belt.\ Astronomical Journal 121, 2804-2814.
  
\item[] Brown, M.~E., and 14 colleagues 2005.\ Keck Observatory Laser
  Guide Star Adaptive Optics Discovery and Characterization of a
  Satellite to the Large Kuiper Belt Object 2003 EL$_{61}$.\ 
  Astrophysical Journal 632, L45-L48.
  
\item[] Brown, M.~E., Schaller, E.~L.\ 2007.\ The Mass of Dwarf Planet
  Eris.\ Science 316, 1585.

\item[] Brown, M.~E., Trujillo, C.~A., Rabinowitz, D.~L.\ 2005.\ 
  Discovery of a Planetary-sized Object in the Scattered Kuiper Belt.\ 
  Astrophysical Journal 635, L97-L100.
  
\item[] Brown, M.~E., Barkume, K.~M., Ragozzine, D., Schaller, E.~L.\ 
  2007. A Collisional Family of Icy Objects in the Kuiper Belt.\ 
  Nature 446, 294-296.
  
\item[] Canup, R.~M.\ 2005.\ A Giant Impact Origin of Pluto-Charon.\ 
  Science 307, 546-550.
  
\item[] Davis, D.~R., Farinella, P.\ 1997.\ Collisional Evolution of
  Edgeworth-Kuiper Belt Objects.\ Icarus 125, 50-60. 

\item[] Dones, L., Weissman, P.~R., Levison, H.~F., Duncan, M.~J.\ 
  2004.\ Oort cloud formation and dynamics.\ Comets II 153-174.

\item[] Doressoundiram, A., Barucci, M.~A., Romon, J., Veillet, C.\ 
  2001.\ Multicolor Photometry of Trans-neptunian Objects.\ Icarus
  154, 277-286.
  
\item[] Duncan, M.~J., Levison, H.~F.\ 1997.\ A scattered comet disk
  and the origin of Jupiter family comets.\ Science 276, 1670-1672.

\item[] Duncan, M.~J., Levison, H.~F., Budd, S.~M.\ 1995.\ The
  Dynamical Structure of the Kuiper Belt.\ Astronomical Journal 110,
  3073-3081.

\item[] Duncan, M., Levison, H., Dones, L.\ 2004.\ Dynamical evolution
  of ecliptic comets.\ Comets II 193-204.
  
\item[] Gladman, B., Kavelaars, J.~J., Petit, J.-M., Morbidelli, A.,
  Holman, M.~J., Loredo, T.\ 2001.\ The Structure of the Kuiper Belt:
  Size Distribution and Radial Extent.\ Astronomical Journal 122,
  1051-1066.

\item[] Gomes, R.~S.\ 2003.\ The origin of the Kuiper Belt
  high-inclination population.\ Icarus 161, 404-418.

\item[] Gomes, R., Levison, H.~F., Tsiganis, K., Morbidelli, A.\ 
  2005.\ Origin of the cataclysmic Late Heavy Bombardment period of
  the terrestrial planets.\ Nature 435, 466-469.
  
\item[] Gomes, R.~S., Fern\'andez, J., Gallardo, T., Brunini,
  A.~2007.\ The Scattered Disk: Origins, Dynamics and End States.\ The
  Solar System Beyond Neptune. 259-273.
  
\item[] Hahn, J.~M., Malhotra, R.\ 2005.\ Neptune's Migration into a
  Stirred-Up Kuiper Belt: A Detailed Comparison of Simulations to
  Observations.\ Astronomical Journal 130, 2392-2414.

\item[] Kenyon, S.~J., Luu, J.~X.\ 1998.\ Accretion in the Early
  Kuiper Belt. I. Coagulation and Velocity Evolution.\ Astronomical
  Journal 115, 2136-2160.
  
\item[] Kenyon, S.~J., Luu, J.~X.\ 1999.\ Accretion in the Early
  Outer Solar System.\ Astrophysical Journal 526, 465-470.

\item[] Kenyon, S.~J., Bromley, B.~C.\ 2002.\ Collisional Cascades in
Planetesimal Disks. I.  Stellar Flybys.\ Astronomical Journal 123,
1757-1775.

\item[] Kenyon, S.~J., Bromley, B.~C.\ 2004.\ The Size Distribution of
  Kuiper Belt Objects.\ Astronomical Journal 128, 1916-1926.
  
\item[] Kenyon, S.~J., Bromley, B.~C., O'Brien, D.~P., Davis, D.~R.\ 
  2008. Formation and Collisional Evolution of Kuiper belt Objects.
  The Solar System Beyond Neptune 293-314.

\item[] Kuchner, M.~J., Brown, M.~E., Holman, M.\ 2002.\ Long-Term
  Dynamics and the Orbital Inclinations of the Classical Kuiper Belt
  Objects.\ Astronomical Journal 124, 1221-1230.
  
\item[] Levison, H.~F., Duncan, M.~J.\ 1994.\ The long-term dynamical
  behavior of short-period comets.\ Icarus 108, 18-36.

\item[] Levison, H.~F., Duncan, M.~J.\ 1997.\ From the Kuiper Belt to
  Jupiter-Family Comets: The Spatial Distribution of Ecliptic Comets.\ 
  Icarus 127, 13-32.
  
\item[] Levison, H.~F., Stern, S.~A.\ 2001.\ On the Size Dependence of
  the Inclination Distribution of the Main Kuiper Belt.\ Astronomical
  Journal 121, 1730-173
  
\item[] Levison, H.~F., Morbidelli, A., Van Laerhoven, C., Gomes, R.,
  Tsiganis, K.~ 2008.  Origin of the structure of the Kuiper Belt
  during a Dynamical Instability in the Orbits of Uranus and Neptune.\ 
  Icarus 196, 258-273.
  
\item[] Malhotra, R.\ 1995.\ The Origin of Pluto's Orbit: Implications
  for the Solar System Beyond Neptune.\ Astronomical Journal 110, 420.

\item[] Morbidelli, A.\ 2007.\ Solar system: Portrait of a suburban
  family.\ Nature 446, 273-274.
  
\item[] \Red{Morbidelli, A., Zappala, V., Moons, M., Cellino, A.,
    Gonczi, R.\ 1995.\ Asteroid families close to mean motion
    resonances: dynamical effects and physical implications.\ Icarus
    118, 132. }
 
\item[] Morbidelli, A., Valsecchi, G.~B.\ 1997.\ NOTE: Neptune
  Scattered Planetesimals Could Have Sculpted the Primordial
  Edgeworth-Kuiper Belt.\ Icarus 128, 464-468.

\item[] Morbidelli, A., Emel'yanenko, V.~V., Levison, H.~F.\ 2004.\ 
  Origin and orbital distribution of the trans-Neptunian scattered
  disc.\ Monthly Notices of the Royal Astronomical Society 355,
  935-940.

\item[] Morbidelli, A., Levison, H.~F., Tsiganis, K., Gomes, R.\ 
  2005.\ Chaotic capture of Jupiter's Trojan asteroids in the early
  Solar System.\ Nature 435, 462-465.
  
\item[] Morbidelli, A., Levison, H.~F., Gomes, R.\ 2007.\ The
  Dynamical Structure of the Kuiper Belt and its Primordial Origin.\ 
  The Kuiper Belt, in press.

\item[] Nagasawa, M., Ida, S.\ 2000.\ Sweeping Secular Resonances in
  the Kuiper Belt Caused by Depletion of the Solar Nebula.\ 
  Astronomical Journal 120, 3311-3322.

\item[] Nesvorn{\'y}, D., Vokrouhlick{\'y}, D., Morbidelli, A.\ 
  2007a.\ Capture of Irregular Satellites during Planetary
  Encounters.\ Astronomical Journal 133, 1962-1976.  

\item[] Nesvorn{\'y}, D., Vokrouhlick{\'y}, D., Bottke, W.~F.,
  Gladman, B., H{\"a}ggstr{\"o}m, T.\ 2007b.\ Express delivery of
  fossil meteorites from the inner asteroid belt to Sweden.\ Icarus
  188, 400-413.
  
\item[] \Green{Noll, K.~S., Grundy, W.~M., Stephens, D.~C., Levison, H.~F.,
  Kern, S.~D.\ 2008.\ Evidence for two populations of classical
  transneptunian objects: The strong inclination dependence of
  classical binaries.\ Icarus 194, 758-768.}

%Noll, K.~S., Grundy, W.~M., Chiang, E.~I., Margot, J.-L.,
%  Kern, S.~ D.\ Binaries in the Kuiper Belt\ The Kuiper Belt, in
%  press.
  
\item[] Petit, J.-M., Holman, M.~J., Gladman, B.~J., Kavelaars, J.~J.,
  Scholl, H., Loredo, T.~J.\ 2006.\ The Kuiper Belt luminosity
  function from $m_{R}=$ 22 to 25.\ Monthly Notices of the Royal
  Astronomical Society 365, 429-438.

\item[] Rabinowitz, D.~L., Barkume, K., Brown, M.~E., Roe, H.,
  Schwartz, M., Tourtellotte, S., Trujillo, C.\ 2006.\ Photometric
  Observations Constraining the Size, Shape, and Albedo of 2003 EL61,
  a Rapidly Rotating, Pluto-sized Object in the Kuiper Belt.\ 
  Astrophysical Journal 639, 1238-1251.
  
\item[] Ragozzine, D., Brown, M.~E.\ 2007.\ Candidate Members and Age
  Estimate of the Family of Kuiper Belt Object 2003 EL61.\ 
  Astronomical Journal 134, 2160-2167.

\item[] Sheppard, S.~S., Trujillo, C.~A.\ 2006.\ A Thick Cloud of
  Neptune Trojans and Their Colors.\ Science 313, 511-514.

\item[] Stern, S.~A.\ 1996.\ On the Collisional Environment, Accretion
  Time Scales, and Architecture of the Massive, Primordial Kuiper
  Belt..\ Astronomical Journal 112, 1203.

\item[] Stern, S.~A., Colwell, J.~E.\ 1997a.\ Accretion in the
  Edgeworth-Kuiper Belt: Forming 100-1000 KM Radius Bodies at 30 AU
  and Beyond.\ Astronomical Journal 114, 841.

\item[] Stern, S.~A., Colwell, J.~E.\ 1997b.\ Collisional Erosion in
  the Primordial Edgeworth-Kuiper Belt and the Generation of the 30-50
  AU Kuiper Gap.\ Astrophysical Journal 490, 879.

\item[] Tegler, S.~C., Romanishin, W.\ 2000.\ Extremely red
  Kuiper-belt objects in near-circular orbits beyond 40 AU.\ Nature
  407, 979-981.

\item[] Thommes, E.~W., Duncan, M.~J., Levison, H.~F.\ 1999.\ The
  formation of Uranus and Neptune in the Jupiter-Saturn region of the
  Solar System.\ Nature 402, 635-638.

\item[] Thommes, E.~W., Duncan, M.~J., Levison, H.~F.\ 2002.\ The
  Formation of Uranus and Neptune among Jupiter and Saturn.\ 
  Astronomical Journal 123, 2862-2883.

\item[] Trujillo, C.~A., Jewitt, D.~C., Luu, J.~X.\ 2000.\ Population
  of the Scattered Kuiper Belt.\ Astrophysical Journal 529, L103-L106.

\item[] Tsiganis, K., Gomes, R., Morbidelli, A., Levison, H.~F.\ 
  2005.\ Origin of the orbital architecture of the giant planets of
  the Solar System.\ Nature 435, 459-461.

\end{itemize}

\clearpage

\end{document}